\documentclass[prd,twocolumn,superscriptaddress]{revtex4}
\usepackage{graphicx}
\usepackage{amssymb}

\providecommand\lesssim{\stackrel{<}{\sim}}

\newcommand\ba{\begin{array}}
\newcommand\ea{\end{array}}
\newcommand\bc{\begin{center}}
\newcommand\ec{\end{center}}
\newcommand\be{\begin{enumerate}}  
\newcommand\ee{\end{enumerate}}  
\newcommand\bi{\begin{itemize}}  
\newcommand\ei{\end{itemize}}  
\newcommand\bd{\begin{description}}  
\newcommand\ed{\end{description}}  
\newcommand\beq{\begin{equation}}  
\newcommand\eeq{\end{equation}}  
\newcommand\beqa{\begin{eqnarray}}  
\newcommand\eeqa{\end{eqnarray}}

\newcommand{\eq}[1]{Eq.\ (\ref{#1})}

\newcommand\mathC{\mkern1mu\raise2.2pt\hbox{$\scriptscriptstyle|$}
                {\mkern-7mu\rm C}}

\def\exl{\raise1pt\hbox{$\scriptstyle<$}}
\def\exr{\raise1pt\hbox{$\,\scriptstyle>$}}

\newcounter{theo}[section]
\renewcommand{\thetheo}{\thesection.\Roman{theo}}

\newcounter{cor}
\renewcommand{\thecor}{\thetheo.\roman{cor}}

\newcounter{exa}

\newcounter{def}[section]
\renewcommand{\thedef}{\thesection.\arabic{def}}

\newcounter{rem}

\begin{document}
\date{\today}
\title{Global, Exact Cosmic Microwave Background Data Analysis Using Gibbs Sampling}
\author{Benjamin D.~Wandelt}
\thanks{Benjamin D.~Wandelt is an NCSA/UIUC Faculty Fellow}
\email{bwandelt@uiuc.edu}
\affiliation{Department of Physics, UIUC, 1110 W Green Street, Urbana, IL 61801}
\affiliation{Department of Astronomy, UIUC, 1002 W Green
Street, Urbana, IL 61801}
\author{David L.~Larson}
\affiliation{Department of Physics, UIUC, 1110 W Green Street, Urbana, IL 61801}
\author{Arun  Lakshminarayanan}
\affiliation{Department of Physics, UIUC, 1110 W Green Street, Urbana, IL 61801}
\
\begin{abstract}
  We describe an efficient and exact method that enables
  global Bayesian analysis of cosmic microwave background (CMB) data.
  The method reveals the joint posterior density (or likelihood for
  flat priors) of the power spectrum $C_\ell$ and the 
  CMB signal. Foregrounds and instrumental parameters can be
  simultaneously inferred from the data. The method allows the
  specification of a wide 
  range of foreground priors.  We explicitly show how to propagate the
  non-Gaussian dependency structure of the $C_\ell$ posterior through
  to the posterior density of the parameters. If desired, the analysis
  can be coupled to theoretical (cosmological) priors and can yield
  the posterior density of cosmological parameter estimates directly
  from the time-ordered data. The method does not hinge on special
  assumptions about the survey geometry or noise properties, etc. It
  is based on a Monte Carlo approach and hence parallelizes trivially.
  No trace or determinant evaluations are necessary. The feasibility
  of this approach rests on the ability to solve
  the  systems of linear equations which arise. These are of the same
  size and computational complexity as the map-making equations.  We describe a
  pre-conditioned conjugate gradient technique that solves this
  problem   and demonstrate in a numerical example that the
  computational time required for each Monte Carlo sample scales as
  $n_p^{3/2}$ with the number of pixels $n_p$. We test
  our method using the COBE-DMR data and explore the non-Gaussian joint
  posterior density of the COBE-DMR $C_\ell$ in several projections.
\end{abstract}
\maketitle

\section{Introduction}
The observation and analysis of cosmic microwave background (CMB)
anisotropies have attracted a great deal of attention in recent years
due to their unique relevance for cosmological theory (see
\cite{review} for a recent review). A slew of observational results
have been published during the last two years\cite{cmbobs}. These were
obtained from maps of the microwave sky at ever increasing sensitivity
and resolution. Since the recent release of the first year WMAP data,
an all-sky microwave survey has been available down to angular scales
of 12 minutes of arc \cite{mapresults}. By the end of the decade the
Planck satellite is expected to generate 1 Terabyte of high
resolution,  high sensitivity all-sky data.

The basic assumption is that the
CMB anisotropy signal and the instrumental noise are Gaussian
and that the signal statistics are isotropic on the sky.
Contact between theory and observation is then best made by extracting the
angular power spectrum $C_\ell$ from the data\cite{BondEfstathiou,Tegmark,BJK}.  Methods for
efficiently estimating the power spectrum have been investigated since
the computational unfeasiblity of using the brute-force approach was
realized \cite{Gorski,BCJK,Borrill}. 
 This effort has yielded two
classes of methods: exact methods, applicable only to two
narrowly defined classes of observational strategies \cite{OSH,WH},
and approximate but more broadly applicable methods
\cite{pseudocl,MASTER,szapudi}\footnote{An exception to this
  classification is a hybrid method which has
been proposed very recently and which combines a maximum likelihood approach on large
scales with an approximate approach on small scales \cite{EfstathiouHybrid}.}.

We will describe here a solution to the problem of inference from
microwave background data which combines the advantages of exact methods with the
practicality of the approximate methods. The computational
cost of our method scales like the best approximate method for the
same experiment, albeit with a larger pre-factor. Power spectrum
estimates and any desired characterization of the (multivariate)
statistical uncertainty in the estimates  can be computed
free from any approximations in the estimator which could lead to sub-optimality or biases. 

The solution we propose is to {\it sample} the power spectrum (as well as
other desired quantities, such as the underlying CMB signal, foregrounds
or the noise properties 
of the instrument) directly from the joint likelihood (or posterior) density given
the data. We can efficiently sample from this multi-million dimensional density
using the Gibbs sampler. This approach obviates the need to 
evaluate the likelihood or its derivatives in order to analyze
CMB data.

Our approach shares certain algorithmic features with the approach
independently discovered in \cite{Jewell} which describes
a maximum likelihood estimator of the power spectrum using Bayesian Monte Carlo
methods. However, our goal from the outset
was to design a method that allows a full exploration of the
multivariate probability density of the power spectrum and the
parameter estimates, given the 
data. 

Our method seamlessly integrates  with parameter
estimation without recourse to semi-analytic Gaussian, offset
log-normal \cite{RadPack}, $\chi^2$ \cite{Bartlett} or hybrid
\cite{WMAPverde} approximation schemes. If desired, theoretical priors
can be applied in the analysis by restricting the space of power spectra
to those which arise from a physical model of the CMB anisotropy.

By design, the sample of power spectra and reconstructed sky maps
will reflect the statistical uncertainty given the
data through the full non-Gaussian statistical dependence structure
of the $C_\ell$ estimates. This information can be propagated
losslessly to the cosmological parameter 
estimates. 

One aspect of our method which is of general interest in astrophysics beyond CMB
analysis is that it generalizes the results 
on globally optimal interpolation, filtering and reconstruction of
noisy and censored data sets in \cite{rybickipress} to
self-consistently include inference of the signal covariance
structure. This defines a generalized Wiener filter that does not need
a priori specification of the signal covariance. A byproduct of our
method is a prescription for ``unbiasing'' the Wiener filter which
clearly reveals the tight relation between Wiener filtering and power spectrum estimation.

Our methods differ from traditional methods of CMB analysis in a fundamental aspect.
Traditional
methods consider the analysis task as a set of steps, each
of which arrives at  intermediate outputs which are then fed as
inputs to the next step in the pipeline. Our approach is a truly
global analysis, in the sense that the statistics of
all the science products are computed jointly, respecting and
exploiting  the full
statistical dependence structure between the various components.

In summary, our method is a Monte Carlo technique which samples power
spectra and other science products from their exact, multivariate a
posteriori probability density, and which does so without explicitly
evaluating it. The result is a detailed characterization of the
statistics of the CMB signal on the sky, reconstructed foregrounds,
the  CMB power spectrum, and the cosmological parameters 
inferred from it with a cost which is proportional to the cost
of a least squares map-making algorithm for the same set of
observations.

In section \ref{notation} we introduce notation and a general
statistical model of CMB observations. Our method is described in
detail in section \ref{method}. In section \ref{cosmicvariance} we
comment on the  perspective our Bayesian approach offers on cosmic
variance. We discuss the numerical and computational techniques used
to implement our method in section \ref{numerics} and apply it to the
COBE-DMR data in section \ref{cobe}. We reflect on where we stand and
conclude with comments on
further work to be done in section \ref{conclusions}.

\section{Model and Notation}
\label{notation}
We begin by defining our model of CMB observations and introduce our
notation. We imagine that the actual CMB sky $s$
is observed with some optical system and according to some observing
strategy encoded in a pointing 
matrix $A$, which maps the signal on the sky into a  collection of
$n_{o}$ time-ordered observations of the sky. 
This results in the ``raw'' data $d$, represented by a vector with $n_{o}$
elements (an $n_o$--vector). Our model of this process is
encoded in the model equation
\begin{equation}
d=A(s+f)+n^{tod},
\label{TODmodel}
\end{equation}
where $n^{tod}$ is a realization of Gaussian instrumental noise added to
the data and $f=\sum_i f_i$ is the sum of a collection of foregrounds (assumed spatially
varying and constant in time). We represent maps on the sky
with $n_p$ resolution elements (pixels) as $n_p$--vectors. Note that
while we do not explicitly consider multi-channel data, the model is easily
generalized to that case by adding a frequency index to $d$, $A$, $n^{tod}$ and
 $f$.

The ``map'' vector $m$ is the least squares estimate of the signal $s+f$
from $d$. Because we assume Gaussian noise with zero mean this is also
the maximum 
likelihood estimate (or maximum a posteriori estimate assuming a
flat prior). It can be found as the solution of the
normal equation
\begin{equation}
A^TN_{tod}^{-1}A m=A^TN_{tod}^{-1}d.
\label{normal}
\end{equation}
Here the matrix $N_{tod}$ is the covariance matrix of the noise in the
time ordered data space $N_{tod}=\langle n^{tod}n^{tod\;T}\rangle $. Then $m=s+f+n$
where $n$ describes the residual noise on the map estimate with
covariance matrix $N=\langle n n^T\rangle =(A^TN_{tod}^{-1}A)^{-1}$.

The cosmological model specifies the signal covariance matrix $S$. For
isotropic theories $S$ is diagonal in the spherical harmonic basis,
with the  special form
$S_{\ell m\ell 'm'}=C_\ell\delta_{\ell\ell'}\delta_{mm'}$. 

In keeping with the majority of the literature in the field, we
restrict our discussion to theories which predict a Gaussian 
CMB signal $s$. It will be convenient to abbreviate Gaussian multivariate
densities as
\begin{equation}
G(m,C)\equiv\frac{1}{\sqrt{\vert 2 \pi C \vert}}\exp\left(\frac12m^TC^{-1}m\right).
\end{equation}

\section{Method}
\label{method}
\subsection{Overview}
For a cleaner exposition of the method, we will ignore
the foregrounds $f$ for now and return to their  inclusion later.
We are trying to explore the {\it a posteriori} density
\begin{equation}
P(C_\ell|m)\propto G(m,S(C_\ell)+N)P(C_\ell)
\label{Pclgivenm}
\end{equation}
where $P(C_\ell)$ is the density encoding prior information on the $C_\ell$.
Up to normalization this can be
obtained by  marginalizing the joint density 
\begin{equation}
P(C_\ell,s,m)= P(m|s)P(s|C_\ell)P(C_\ell)
\label{Pjoint}
\end{equation}
over the signal $s$. Setting $P(C_\ell)=const$  makes this analysis
equivalent to an exact frequentist likelihood analysis. We will
discuss other choices of prior later.

Traditionally, the approach to exploring the {\it a
posteriori} 
density has been to define an estimator, such as the least squares
quadratic (LSQ) estimator \cite{Tegmark} or the maximum
likelihood (ML) estimator \cite{BJK}.  Then
some measure of uncertainty in the values of this estimator was
defined, for instance by approximating the shape of $P(C_\ell|m)$ around
the maximum by a multivariate Gaussian and evaluating elements of the
curvature matrix at the extremum. 

Evaluating the LSQ or ML estimators is a very
costly operation, taking $O(n_p^3)$ operations\footnote{See however
  \cite{pen} for fast numerical techniques that were applied
  successfully to weak lensing data.}. In general, evaluating 
the curvature matrix is
even more costly because it has $O(n_p)$ elements
{\it each of which} requires  $O(n_p^3)$ operations, making the
overall operation count of order   $O(n_p^4)$.
In addition a Gaussian approximation fails at low $\ell $
where the small number of degrees of freedom makes the posterior
significantly non-Gaussian, and also at high $\ell $ in the regime of 
small signal-to-noise ($S/N\lesssim 1$). 

Instead, we propose to sample parameter values $C_\ell$ from the
posterior directly. There is no known way to directly sample from
Eq.\ref{Pclgivenm}, but if a way can be found to sample $s$ and $C_\ell$
from the joint distribution \eq{Pjoint} then the $C_\ell$ taken by
themselves are exact samples from the marginalized distribution.

At first, sampling from the joint distribution seems even less feasible. But
powerful theorems can be proved \cite{tanner} that show that if it is possible to
sample from the {\it conditional distributions} $P(s|C_\ell,m)$ and
$P(C_\ell|s,m)\propto P(C_\ell|s)$  then one can sample from the joint
distribution in an iterative fashion. Begin with some starting guess
$C_\ell^0$. Then iterate the following equations
\begin{equation}
s^{i+1}\leftarrow P(s|C_\ell^{i},m)
\label{signaldraw}
\end{equation}
\begin{equation}
C_\ell^{i+1}\leftarrow P(C_\ell|s^{i+1})
\end{equation}
then after some ``burn-in'' the $(C_\ell^i,s^i)$ converge to being samples from the
joint distribution \eq{Pjoint}.
This technique of sampling from the joint distribution is called the {\it
Gibbs sampler}. 

\subsection{Sampling Techniques}

To implement these ideas one needs the forms of the conditional
densities and  recipes for sampling from these
distributions. These follow now. 

The conditional density of the signal given the most recent $C_\ell$
sample is just a multivariate Gaussian
\begin{equation}
P(s|C_\ell^{i},m)\propto G\left(S^i(S^i+N)^{-1}m, ((S^i)^{-1}+N^{-1})^{-1}\right),
\label{PsgivenCl}
\end{equation}
where $S^i\equiv S(C_\ell^i)$. This will be recognized as the posterior
density of the {\it Wiener Filter} given the most recent power
spectrum estimate. 

The density for the power spectrum coefficients $C_\ell$
factorizes due to the special form of $S$. 
\begin{equation}
P(C_\ell|s^i)\propto P(C_\ell)\prod_l
\frac{1}{\sqrt{C_\ell^{2\ell +1}}}\exp\left(-\frac1{2C_\ell}\sum_{m=-l}^{+l}
\vert s^i_{\ell m}\vert^2\right)
\label{PClgivens}
\end{equation}
The $s^i_{\ell m}$ are the spherical harmonic coefficients of $s^i$.
This density is known as the inverse Gamma distribution of order
$2\ell -1$. This result has interesting implications for cosmic
variance in this Bayesian framework, which we will discuss below.

To sample from \eq{PsgivenCl} we need to generate a Gaussian
variate with the given mean and covariance. A numerically convenient
choice (see section \ref{numerics}) of the equation for the  mean $x$ is 
\begin{equation}
(1+S^{i\;\frac12}N^{-1}S^{i\;\frac12})S^{i\;-\frac12}x= S^{i\;\frac12}N^{-1}m.
\label{Wienermean}
\end{equation}
In fact it is easier to solve for $z=S^{i\;-\frac12}x$ and to then
solve for x trivially.
Note from its definition above that $N^{-1}x$ is easier to compute
than $Nx$. If $N_{tod}$ is circulant (stationary noise) or
block-circulant (a popular choice for non-stationary noise),
$N_{tod}^{-1}x$ can be computed 
using FFTs. If $N$ is diagonal to very good accuracy then computing
$N^{-1}$ easy. We will drop the $i$ superscript in what follows.

We chose to write the equation in terms of the map made from
the data. It easy to see from \eq{normal} and from
$N=(A^TN_{tod}^{-1}A)^{-1}$ that writing \eq{Wienermean}
in terms of the TOD saves some computations:
$N^{-1}m=A^TN_{tod}^{-1}d$. This replacement can be made throughout in the
equations that follow in the remainder of this paper.

Then we need to add a fluctuation term to this mean to get a random
variate. This is non-trivial, because we need to simulate a map with
covariance $(S^{-1}+N^{-1})^{-1}$ without being able to compute square roots
of this matrix. We can, however, compute the square root of $S$
because it is
diagonal in spherical harmonic space and we can compute
$N^{-\frac12}\equiv A^TN_{tod}^{\frac12}$ by using FFTs on the time-ordered
data.  This leads to the following solution: generate two  
p-vectors $\xi$ and $\chi$ of independent Gaussian random variates, with
zero mean and unit variance (these are called {\it normal}
variates). Then solve the linear set of equations  
\begin{equation} 
\left(1+S^{\frac12}N^{-1}S^{\frac12}\right)S^{-\frac12}y=\xi+S^{\frac12}N^{-\frac12}\chi
\label{Wienerfluct}
\end{equation}
for y. It is easy to verify that this does give  the right covariance by
computing $\langle yy^T\rangle $. The final result is then
\begin{equation}
s^{i+1}=x+y,
\end{equation}
where we have re-introduced the superscript.

Note that each $s$ is a perfect pure signal sky (up to the assumed
band-limit) with covariance $S$. While $x$ is the Wiener filter,
whose power spectrum would be a biased estimator of the $C_\ell$, $s$
is ``unbiased''. The addition of the fluctuating term $y$  has
replaced filtered noise fluctuations with synthetic signal fluctuations. 

Drawing the $C_\ell^{i+1}$ from the inverse Gamma distribution,
\eq{PClgivens}, is very simple. For each $\ell$, compute 
$\sigma_\ell=\sum_{m=-\ell}^{+\ell}  
\vert s^i_{\ell m}\vert^2$ and generate a $(2\ell-1)$-vector $\rho_\ell$  
of Gaussian random variates with zero mean and unit variance. Then
\begin{equation}
C_\ell^{i+1}= \frac{\sigma_\ell}{\vert\rho_\ell\vert^2},
\label{sampleCl}
\end{equation}
where the denominator is the square norm of $\rho_\ell$.

\subsection{Foregrounds} 
%
Traditionally, regions on the sky are excised if the residual error
after foreground subtraction is large. However, modeling the signal on
the remainder of the sky after foreground cuts complicates the
structure of the signal covariance matrix $S$. Instead, we choose to
model foregrounds as an additional component in the model equation, as
shown in \eq{TODmodel}. Then the joint density in \eq{Pjoint} becomes
\begin{equation}
P(C_\ell,s,\{f_j\},d)= P(d|s,\{f_j\})P(s|C_\ell)P(C_\ell)\prod_kP(f_k)
\label{Pjoint+}
\end{equation}
where each $P(f_k)$ contains prior information about the $k$th
foreground. 

Following the Gibbs sampler approach we draw from the foreground
components given the data. We group different logically separate 
foregrounds by adding in  additional steps in the sampling
chain
\begin{eqnarray}
{\rm ~for~every~}j:~f_j^{i+1}&\leftarrow& P(f_j|C_\ell^{i},s^i,\{f_{k<
  j}\}^{i+1},\{f_{k> j}\}^{i})\nonumber\\ 
s^{i+1}&\leftarrow& P(s|C_\ell^{i},\{f_j\}^{i+1})\label{Gibbs}\\
C_\ell^{i+1}&\leftarrow& P(C_\ell|s^{i+1})\nonumber
\end{eqnarray}
Where appropriate, different foreground components may be grouped 
together into one $f_j$. The algorithm to sample from the conditional 
foreground densities is analogous to the signal sampling algorithm
described in the previous subsection. We will return to algorithmic
issues after discussing the foreground prior $\prod_jP(f_j)$.

How do we specify the foreground prior? For instance, we might want to
be completely insensitive to 
certain foreground terms $\{f_i\}$. This would mean
setting $P(f)=G(f,FF^T)$, where $FF^T\equiv\sigma_f^2 \sum_i
f_if_i^T$ and  each vector $f_i$ represents a foreground contribution we would
like to project out. The matrix $F$ is just constructed by columns
from the $f_i$. The variance $\sigma_f^2$ is numerically ``infinite'',
i.e. large compared to
any other noise source. This specifies maximal ignorance about the
amplitude of this foreground component. As an example, $f_i$ could be the
monopole and the three dipole 
components. Or, if the foreground contribution 
in a pixel $j$ was completely unknown, the $f_i= 1_j$ where $ 1_j$ is the vector
representing the map which is zero everywhere except in the pixel
$j$. Essentially any spatial template to which we want the power
spectrum estimation to be insensitive can be added in here, and they
can be grouped in computationally convenient ways in \eq{Gibbs}.

It is important to note that even though we may have specified
``infinite'' variance in our prior, the foreground samples produced will be
constrained by the data and hence will be informative about the level of the
foreground contribution supported by the data. For example, the sample
of the three dipole components generated during the iteration of
\eq{Gibbs} in the  example above would be informative about the
direction and amplitude of 
the CMB dipole, and could be used to calibrate the experiment. 

Different choices for the foreground prior $P(f)$ are possible. It could include
information on foreground templates as
well as a specification of our uncertainty in  these 
templates. For example if the template is $\bar{f}$ and our uncertainty
could be described by a Gaussian centered on $\bar{f}$ with covariance 
$FF^T$ then $P(f)=G(f-\bar{f},FF^T)$. One way to specify $\bar{f}$ 
and $FF^T$  would be to simulate a set of possible theoretical
foreground models $f_i$ with weights $w_i$, such that $\sum w_i=1$,
and to then set $\bar{f}=\sum w_i f_i$ and $FF^T\equiv\sum_iw_i
(f_i-\bar{f})(f_i-\bar{f})^T$. 

Note that the assumption of a Gaussian prior
$P(f)$ only assumes that our ignorance of the foreground contribution can
be expressed through a Gaussian covariance structure---the foregrounds
are not assumed to have Gaussian statistics. Non-Gaussianity can be
explicitly assumed by choosing a non-Gaussian template $\bar{f}$.
For the case of
multi-frequency data, $P(f)$ could encode what is known about the dependence of certain
physical foreground components on the frequency.

Returning to the mechanics of sampling \eq{Gibbs} we write ${\cal
  F}_j=F_jF^T_j$, and solve at the $j$th step
\begin{equation}
({\cal F}_j+{\cal F}_j N^{-1}{\cal F}_j)x_j= \bar{f}_j+{\cal F}_j
  N^{-1}(m-s-\sum_{k\ne j} f_j),
\label{WienermeanFG}
\end{equation}
and 
\begin{equation}
\left({\cal F}_j+{\cal F}_jN^{-1}{\cal F}_j\right)y_j=F_j\xi+{\cal F}_jN^{-\frac12}\chi.
\label{WienerfluctFG}
\end{equation}
Then $f_j={\cal F}_j(x_j+y_j)$. Since ${\cal F}_j$ may not be full
rank in $n_p$ dimensions, the equations here may be understood as shorthand for the
projected equations in the subspace on which ${\cal F}_j$
has full rank.

Note that when foregrounds are considered, the $m$ on the right hand side of
\eq{Wienermean} has to be replaced with $(m-\sum_{j} f_j)$.

In  special cases it may be desirable to perform the
marginalization over $f$ analytically. Appendix \ref{marginalization} gives
techniques for doing so.

\subsection{Noise model}
It is straightforward to
extend our methods to include estimation of the noise covariance from
the data themselves. In the case that $N_{tod}$ is non-stationary and
block-diagonal with 
circulant blocks (the standard assumption in CMB analysis), we
can easily add a sampling step symbolically written as
\begin{equation}
N_j^{i+1}\leftarrow P(N|\{N_{k\neq j}\}^{i},s,C_\ell,\{f_j\})=P(N|s,\{f_j\}).
\end{equation} 
The second equality expresses two facts: (1) for a block diagonal noise
matrix the conditional density of one block does not depend on the
other blocks and (2)  $N$ is conditionally independent of the $C_\ell$ given
$s$; that is  given $s$ the $C_\ell$ do not 
add more information about the $N$.

In practice, the noise model would assume smoothness of the noise
power spectra. If we write $N_{jk}$ for the $k$th band power spectral coefficient of the
$j$th block of the noise covariance of the TOD simply involves computing the FFT of  the
$j$th segment of $d-A(s+\sum f)$, adding the power in bands of width
$d$ and then sampling $N_{jk}$ from the inverse Gamma distributions of
order $d-2$. 

More general noise models can be implemented. We will explore the
effect of more sophisticated modeling of non-stationary noise in
future work.

\subsection{Parameter estimation}
Currently power spectrum estimation algorithms rely on approximate
representations of the posterior density $P(C_\ell|d)$ \footnote{We write
  $P(C_\ell|d)$ as shorthand for the multivariate posterior density, a
  function of $\{C_\ell:\;\ell=1,\dots,\ell_{max}\}$. }, for
example in terms of multivariate Gaussian, shifted log-normal or hybrid
representations. These approximations have to be fitted to sets of
Monte Carlo simulations \cite{WMAPverde}. Since they take simple
analytical forms 
they  can only be expected to be
accurate near the peak of the posterior density. In order to
faithfully propagate all the information 
in the $C_\ell$ estimates through to the parameter estimation step,
efficient ways must be found to accurately represent and communicate
$P(C_\ell|d)$. 

The Bayesian estimation technique described in this paper provides a
natural answer to this problem. The method generates a set of samples
from $P(C_\ell|d)$ which can simply be published
electronically. Meaningful summaries of the properties of $P(C_\ell|d)$
can all be calculated arbitrarily exactly, given a sufficient number of
samples.

The disadvantage of using this sample set for parameter estimation is
that it does not lend itself easily to computing a numerical
probability density for a theoretical $C_\ell$ power spectrum computed
from a set of cosmological parameters $\theta$.

However, a fortunate circumstance solves the problem of finding an
arbitrarily exact numerical representation of $P(C_\ell|d)$. At each
iteration of the Gibbs sampler the $C_\ell$ are drawn from $P(C_\ell|s)$
which is in fact $P(C_\ell|\sigma_\ell)$   where $\sigma_\ell=\sum_m
|s_{\ell m}|^2$. 
We can therefore write
\begin{eqnarray}
P(C_\ell|d)&=&\int ds P(C_\ell,s|d)=\int ds P(C_\ell|s)P(s|d)\nonumber\\
&=&\int D\sigma_\ell P(C_\ell|\sigma_\ell)P(\sigma_\ell|d)\approx\frac{1}{n_G}\sum_i  P(C_\ell|\sigma^i_\ell).\nonumber\\\label{BlackwellRao}
\end{eqnarray}
The sum  (where the index runs over $n_G$ Gibbs samples) becomes
an arbitrarily exact approximation to the integral as the number of
samples increases. It is called the {\it Blackwell-Rao} estimator for
the density and can be shown to be superior to binned
representations. This sum yields a numerical representation of the
posterior density of the power spectrum given the signal samples. All
the information about $P(C_\ell|d)$ is contained in the  ${\sigma^i_\ell}$,
which generate a data set of size
$O(\ell_{max}n_G)$.

It is noteworthy that in the limit of perfect data, using
\eq{BlackwellRao} returns the {\it exact} posterior density after only {\it one} iteration of the Gibbs 
sampling algorithm.

In addition to being a faithful
representation of $P(C_\ell|d)$
it is also a computationally efficient representation. Evaluating the
Gaussian or the shifted log-normal 
approximations to $P(C_\ell|d)$ takes $O(\ell_{max}^3)$ operations, while our
approach requires only $O(\ell_{max}n_G)$ operations.   
Note also that any moments of  $P(C_\ell|d)$ can be calculated through
\begin{equation}
\langle C_\ell C_{\ell '}\dots C_{\ell ''}\rangle \vert_{P(C_\ell|d)}\approx\frac{1}{n_G}\sum_i
\langle C_\ell C_{\ell '}\dots
C_{\ell ''}\rangle \vert_{P(C_\ell|\sigma^i_\ell) }.
\end{equation}
This is a far more efficient representation than would be afforded by
a Monte Carlo sample of a pseudo-$C_\ell$ estimator since each of the
terms on the right hand side can be computed analytically.

Another feature of this framework is that is possible to include cosmological
parameter estimation in the joint analysis of  the data.
If we assume a class of theoretical models, we
can solve the  
estimation problem of power spectrum and cosmological parameters
concurrently. The assumption of such a class of models which amounts
to choosing 
a prior for the power spectra which  excludes spectra
that could not 
possibly be the result from a solution of the Boltzmann equation for
any combination of the parameters about which we wish to make inferences.

With such an assumed class of models the relationship between
$C_\ell$ and the cosmological parameters $\theta$ is a non-stochastic
one, $C_\ell=C_\ell(\theta)$,
and $P(C_\ell|\theta)$ is a delta function. We can integrate out this
delta function in the posterior and then  obtain the conditional
density for sampling the cosmological parameters given the data. This
procedure results in removal of the $C_\ell$ sampling step and the
addition of  the following step to the list in \eq{Gibbs}:
\begin{equation}
\theta^{i+1}\leftarrow P(C_\ell(\theta)|s^{i+1}).
\end{equation}
Here $P(C_\ell(\theta)|s^{i+1})$ denotes the inverse Gamma distribution,
\eq{PClgivens}, 
and $C_\ell(\theta)$ is defined through cosmological theory. Instead of
sampling from the $\ell _{max}$ power spectrum 
coefficients given the $\sigma_\ell$ we sample from $\theta$ assuming
that we just measured the $\sigma_\ell$ on a perfect signal sky (the last
draw). In practice, that can be achieved by running a  Markov
Chain using the Metropolis Hastings algorithm until one
independent $\theta$ sample is produced. 

If we believe strongly in the theoretical framework, using this prior
information is desirable: it reduces the number of
parameters in the problem and therefore improves the signal
and hence also the foreground reconstruction from the data. The set of $C_\ell$
for the draws of $\theta$ represents stochastically what is known about the
theoretical power spectrum. This method defines an optimal non-linear
filter which returns the best power spectrum and a characterization of
the error while including physical constraints on the analysis (for
example the smoothness of the 
$C_\ell$ which is related to the natural frequency of oscillations
modes in the primordial plasma).  

However, just as we are interested in making maps from the data
without inputting information about the foregrounds and the
statistical properties (e.g. isotropy) of the CMB, we are also interested in
the model independent power spectrum constraints.

\begin{figure}
\includegraphics[  width=.52\textwidth,
  keepaspectratio,
  angle=0]{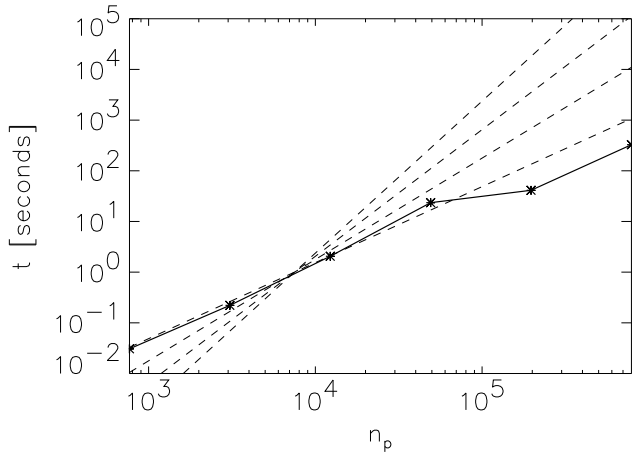}
\caption{Computing time averaged over 30 iterations of the Gibbs
  sampler  required for solving \eq{Wienermean} and
  \eq{Wienerfluct} as a function of the number of pixels in the
  map. These timings are for a single AthlonXP 1800+ CPU. Solid line:
  actual timings. Dashed lines show $n_p^x$ for 
  $x\in\{3,5/2,2,3/2\}$ from the top to the bottom on the right side
  of the figure.
}
\label{scaling}
\end{figure}

\begin{figure}
\includegraphics[  width=.49\textwidth,
  keepaspectratio,
  angle=0,origin=lB]{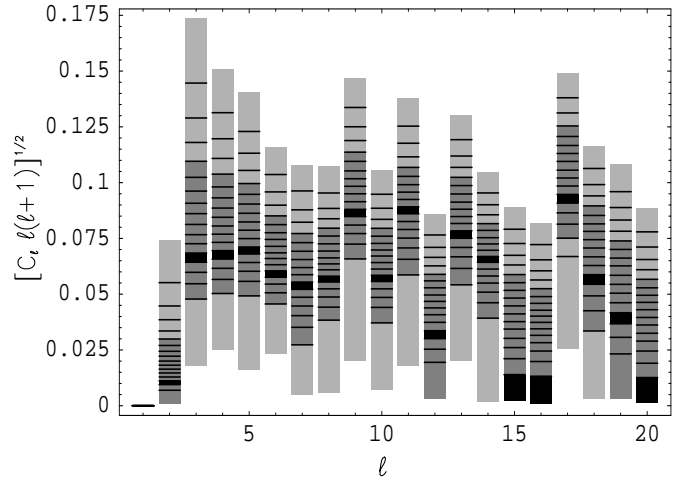}
\caption{The COBE-DMR power spectrum. The vertical bands
  display the marginalized densities at each $\ell$. Horizontal bars mark
  off bins of constant probability. These bins are assigned their
  color in  $C_\ell$ space 
  and then projected into the diagram. The bin with the highest
  probability density is shown in black. The dark and light shaded
  regions are the 1-$\sigma$  and 2-$\sigma$ highest posterior density
  regions, respectively. The $C_\ell$ are measured in units  $mK^2$ in
  this and all subsequent figures.}
\label{powspec}
\end{figure}

\begin{figure*}
\includegraphics[  width=.8\textwidth,
  keepaspectratio,
  angle=0,origin=lB]{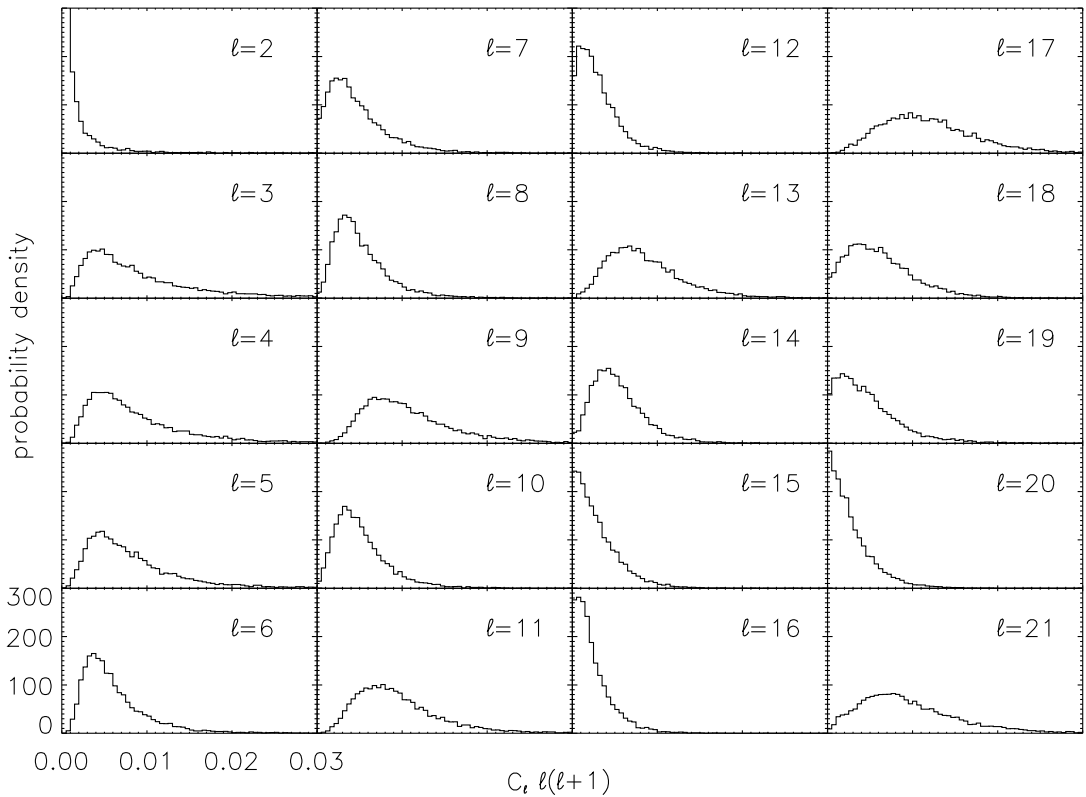}
\caption{Marginalized posterior densities for each individual $C_\ell$
  from the COBE-DMR data. At each $\ell $ the fluctuations in the $C_\ell$ at
  all other $\ell $ were integrated out. The axis ranges are the same
  for all panels.}
\label{posterior1D}
\end{figure*}

\section{Cosmic variance}
\label{cosmicvariance}
In \eq{PClgivens} we have written down the conditional posterior
$P(C_\ell|s)$. This encodes what we know about the $C_\ell$ if we have
perfect knowledge of the signal on the sky. The full
posterior distribution $P(C_\ell|d)$ would reduce to this
if we had perfect (noiseless, all-sky) data. 

As shown in \eq{PClgivens} the $C_\ell$ only depend on
the data through $\sigma_\ell=\sum_{m=-\ell}^{+\ell}  
\vert s^i_{\ell m}\vert^2$. These
$\sigma_\ell$ have a physical interpretation. They measure the actual
fluctuation power on our sky. Therefore, if we had perfect data it
would be possible to measure the $\sigma_l$ with zero
variance. 

The residual uncertainty in $C_\ell$ even for  perfect data is a well
known fact, known as cosmic variance. It is the consequence of having
only one sky at our disposal, which means that there are a limited
number of degrees of freedom for each $C_\ell$. Hence we cannot
measure the $C_\ell$ arbitrarily precisely.

In this Bayesian treatment  the functional form
of the conditional posterior density may be unexpected. In the
frequentist approach where the 
true underlying theory (i.e.~the $C_\ell$) is thought of as fixed and the
data as random, the variances on our sky $\sigma_\ell=\sum_m |s_{\ell m}|^2$ are thought of as
$\chi^2$ variates with $2\ell +1$ degrees of freedom.

From a Bayesian perspective the data is fixed and our knowledge of the
underlying theory is uncertain---so our knowledge about the $C_\ell$ is
encoded in the inverse Gamma distributions $(2\ell -1)$, \eq{PClgivens}.

The mean and variance of the inverse Gamma distribution of order $d$ are 
\begin{equation}
\langle C_\ell\rangle =\frac{\sigma_\ell}{d-2 }\quad d>2,
\end{equation}
and
\begin{equation}
 \langle \Delta C_\ell^2\rangle =\frac{2\,\sigma_\ell^2}{\left( d-4 \right) \,{\left( d-2
 \right) }^2}\quad d>4. 
\end{equation}
For the case of a flat prior $P(C_\ell)=const$ we obtain $d=2\ell -1$. In
this case the variance only 
becomes finite  for $\ell >2$. This is a result 
of having chosen a flat prior for a variance measurement. There are in
fact no arguments for doing so ---when measuring a variance
(which is a scale parameter) a  flat prior does
not correspond to maximal ignorance. 
 
The Jeffrey ignorance prior \footnote{Jeffrey's ignorance prior is
  constructed by requiring that the probability measure $P(C_\ell)dC_\ell$ be invariant
  under transformations which leave our prior knowledge about the parameter
  invariant. In the case of power spectrum estimation (which is
  essentially a variance measurement) we are estimating a positive
  semi-definite scale  parameter $C_\ell$. Our {\it a priori} ignorance
  about the scale implies that the measure must be invariant under
  scale transformations. This is uniquely satisfied if $P(C_\ell)\propto
  1/C_\ell$.} for this case  is $P(C_\ell)=1/C_\ell$. This 
would lead to $d=2\ell +1$ and finite variance for $\ell >1$. In this case
$\langle C_\ell\rangle =\frac{\sigma_\ell}{2\ell -1}$ and $\langle \Delta
C_\ell^2\rangle =\frac{2\,\sigma_\ell^2}{\left( 2\ell -3 \right) \,{\left( 2\ell -1 \right)
  }^2}$.

In order to obtain the frequentist expectation
$\langle C_\ell\rangle =\frac{\sigma_\ell}{2\ell +1}$ the prior $P(C_\ell)=1/C_\ell^2$ would have to
be used. In this case we still obtain a variance for the estimator which
is larger by a factor  $\frac{2\ell +1}{2\ell -1}$ than the frequentist
chi-square variance \cite{knox1995}.  So the mean of  $P(C_\ell|d)$ is an
unbiased estimator of  $C_\ell$ for perfect data and hence has the same
expectation as the maximum likelihood estimator \cite{Jewell}.

These considerations are potentially relevant
to the discussion about the statistical significance of the low $\ell $
$C_\ell$ estimates  in the WMAP data in the Bayesian approach
(e.g., \cite{EfstathiouQP} and references therein). We will
explore this issue in more detail in a future publication.

\begin{figure*}
\includegraphics[width=.75\textwidth,
  keepaspectratio,
  angle=0]{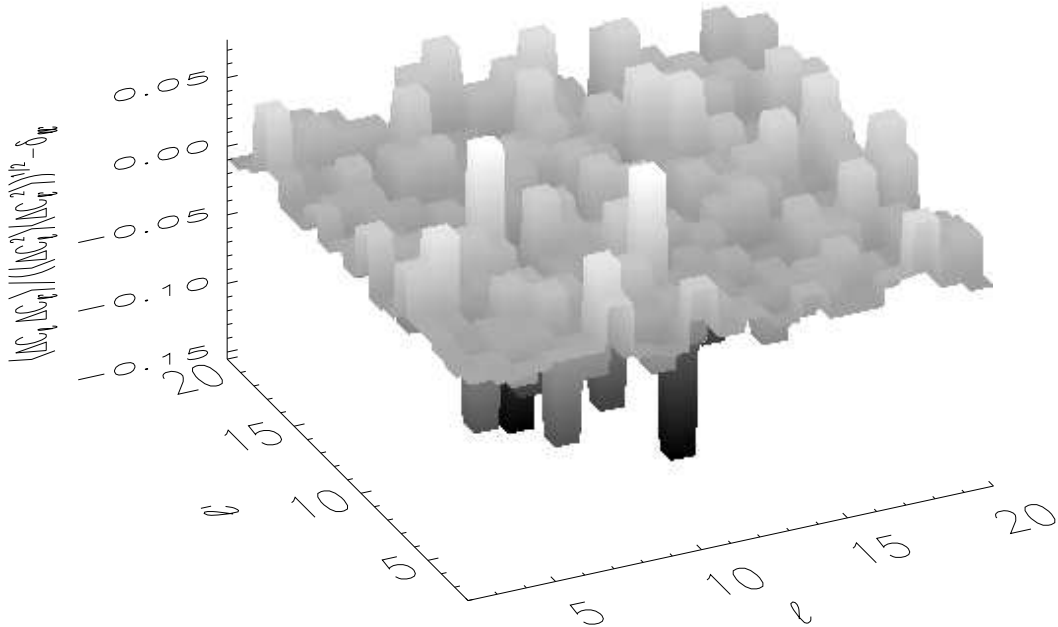}
\caption{Correlation matrix of $C_\ell$ estimates from the COBE-DMR
  data. The diagonal components have been set to zero to enhance the contrast of
  the off-diagonal components. The surface is shaded according to
  height. We see that correlations between the power spectrum
  estimates vary between $8$\% correlation at $(\ell,\ell')=(6,10)$ and
  $15$\% anti-correlation at $(\ell,\ell')=(8,12)$. See Figure \ref{posterior2D}.} 
\label{qMatrix}
\end{figure*}

\begin{figure*}
\includegraphics[  width=\textwidth,
  keepaspectratio,
  angle=0]{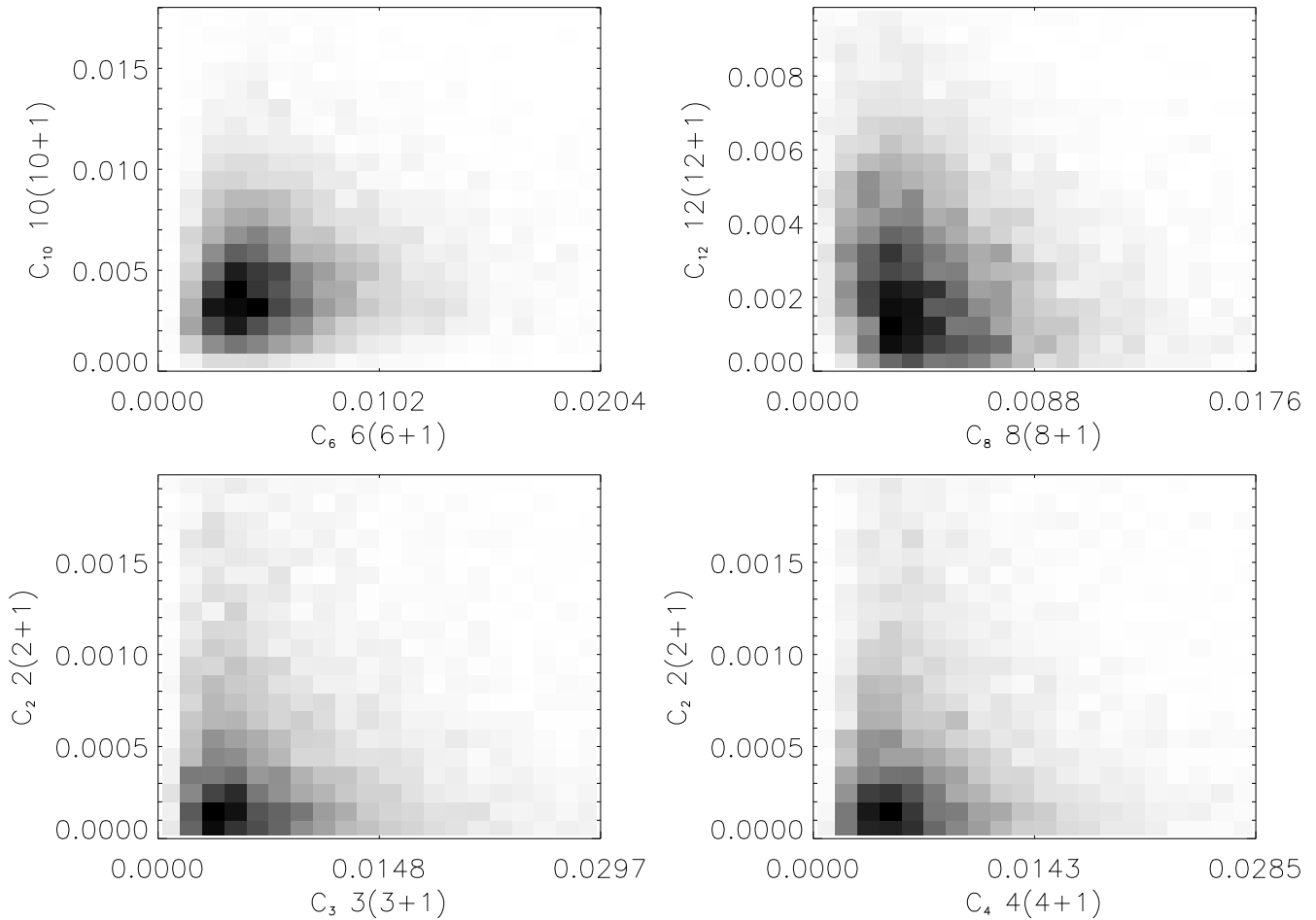}
\caption{2-D marginalized  posterior densities. Each plot shows the
  full joint posterior of the data, integrated over all dimensions
  except for the two shown. From bottom left 
  anti-clockwise: $P(C_2,C_3)$, $P(C_2,C_4)$, $P(C_8,C_{12})$, and
  $P(C_6,C_{10})$.  The latter two were chosen because these $C_\ell$
  pairs  were maximally anti-correlated and correlated, respectively.
}
\label{posterior2D}
\end{figure*}

\begin{figure}
\includegraphics[  width=.45\textwidth,
  keepaspectratio,
  angle=0]{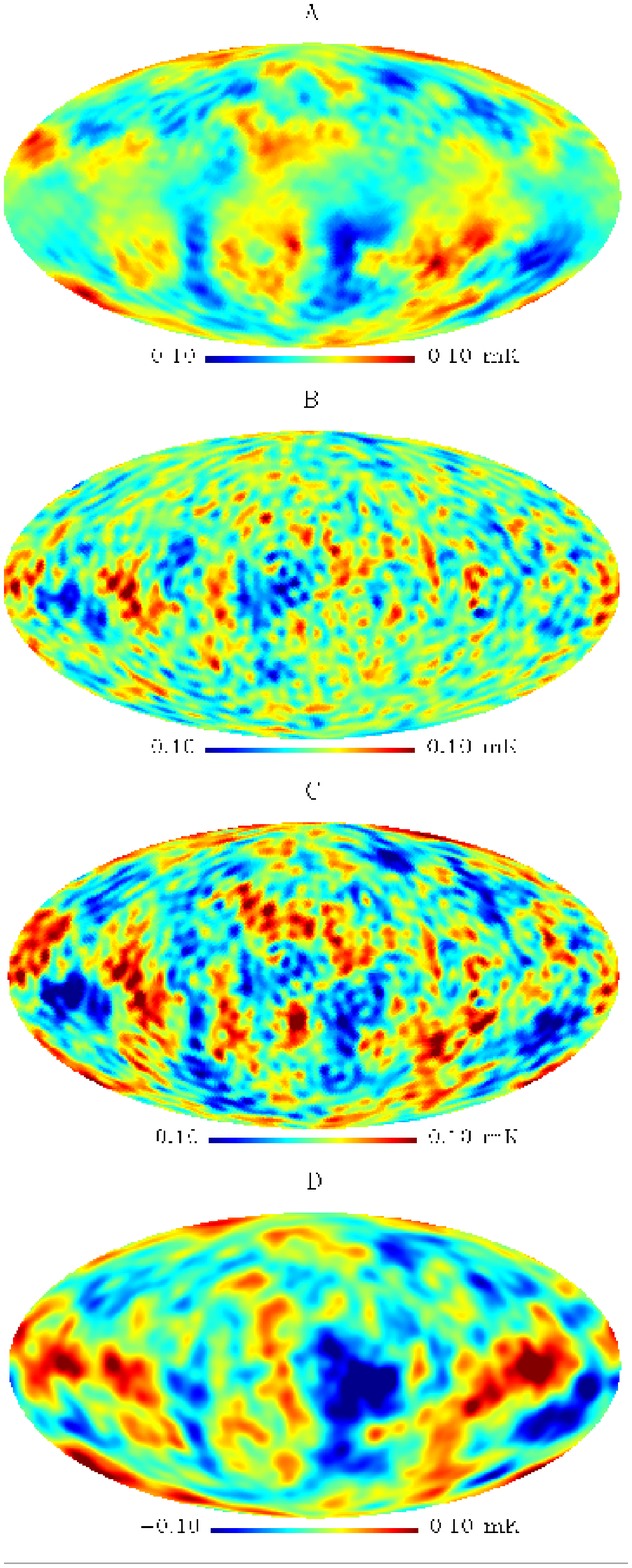}
\caption{Reconstructed signal maps in Galactic coordinates. A: The  signal component of the
  COBE-DMR data marginalized over the power spectrum:   $\langle 
  x\rangle\vert_{P(x|m)}$. This is a  generalized Wiener filter which does not require knowing the signal
  covariance a priori. B: The
  solution $y$ of \eq{Wienerfluct} at one Gibbs iteration. C: The
  sample pure signal sky $s=x+y$ at the same iteration (band-limited
  at $\ell_{max}=50$). D: The WMAP internal linear
  combination map smoothed to an FWHM of 5 degrees. The corresponding
  features in parts A and D are clearly visible. Note that in this map
  low galactic latitudes are not masked, which leads to some artifacts
  that are not visible in the masked COBE-DMR data. 
}
\label{maps}
\end{figure}

\section{Computational Considerations}
\label{numerics}
The computationally most demanding part of implementing this method is
solving \eq{Wienermean} and \eq{Wienerfluct} at each iteration of the
Gibbs sampler. Each of these is a linear system of equations of the
form $Mv=w$, where $M=(1+S^{\frac12}N^{-1}S^{\frac12})$. It should be
noted that these systems are of the same size as
the map-making equation, \eq{normal}.  Maps also have to be made for
approximate estimators. Therefore we expect the computational
complexity of the Gibbs sampler to be no worse than the computational
complexity of an approximate method.

For large $n_p$ $(n_p>10^5)$ on the largest supercomputers available at
the time of writing), direct solution of either of these equations becomes 
infeasible, because neither of them are sparse. This means  the operation count scales as $n_p^3$ and
because the memory requirements for storing the coefficient matrices
scales as $n_p^2$. Therefore large systems of this type are usually
solved using iterative techniques, such as the conjugate gradient (CG)
technique \cite{numrec}. The memory savings can be very large: the
components of $M$ do not have to be stored as long as matrix vector
products $Mv$ can be computed somehow.  In terms of CPU time,
iterative techniques outperform direct 
techniques if either $Mv$ can be computed in less than $n_p^2$
operations or the number 
of iterations required to converge to a solution of sufficient
accuracy is much less than $n_p$. 

We chose to
write  \eq{Wienermean} and \eq{Wienerfluct} in a form which satisfies
all of theses requirements. The memory required is of order
$n_p$ since we never need to  store the
components of the coefficient matrix. 

The action of any power of $S$ on a
vector can be computed in much less than $n_p^2$ operations using  spherical
harmonic transforms (or FFTs in the flat sky approximation). The
action of $N^{-1}=A^TN^{-1}_{tod}A$ on a vector is generally easier to
compute than the 
action of $N$ on a vector. As long as noise correlations can be modeled
in a simple way in the time-domain (e.g. as piecewise 
stationary) the time required for applying $N^{-1}$ to a vector is
similar to that required for a forward simulation of the data. 

The number of CG iterations until convergence can be reduced far
below the theoretical maximum $n_p$ if $M$ is
nearly proportional to the unit matrix. This goal can be approached by
finding an approximate inverse for $M$, a preconditioner. 

If $N^{-1}$ were diagonal in the spherical harmonic basis,  $M$ would 
be, too.   
Therefore, as long as this is approximately
true on scales where $S\gg N$, a good preconditioner for this system would be the
inverse of the  diagonal part of $M$ in the spherical harmonic
basis. These are easy to compute if we approximate the diagonal
components of $N^{-1}$ by
counting the number of $TOD$ samples in each pixel and weighting by
the current noise temperature of the detector.
Due to the way \eq{Wienermean} and \eq{Wienerfluct} have been
written, the structure of $N^{-1}$ in the noise dominated regime does
not matter, since if $S\ll N$, $M\approx 1$. 

This preconditioner can be computed in $O(n_p^\frac32)$
operations. Figure \ref{scaling} shows the results of a timing
study for simulated data sets of varying size with WMAP-like scanning strategy and
uncorrelated noise. The preconditioner performs well. The number of iterations
does not increase with problem size over three orders of magnitude in
$n_p$ and  the computing time is 
is dominated by the spherical harmonic transforms.

\section{Application to the COBE data}
\label{cobe}

In order to test  our method we applied to the well-studied COBE-DMR
data. The exact maximum likelihood estimator
\cite{gorskidmr,gorskimoriond,BJK} and the least square
quadratic estimator \cite{Tegmark} have been computed for this data
set. However, even for this small data set, the marginalized
probability densities of each individual $C_\ell$, or the joint posterior
density of pairs of $C_\ell$ have not been computed because doing so
would require numerical integration over $\sim 20$ dimensions. We will
show these densities here for the first time.

The COBE-DMR data \cite{COBE} is published in the quadcube data structure, at a resolution 
which has 6144 pixels on the sphere. We use a noise-weighted average
of the 53GHz and 90GHz maps.  Because much of our code was
already written for a HealPix data structure, we put the COBE data
into a HealPix pixelization at resolution $n_{side}=64$ with $49152$ pixels.
HealPix pixels whose centers lie within the same quadcube pixel get the
same data (temperature) value.

Because the noise is completely correlated between sets of HealPix pixels
in the same quadcube pixel, the noise covariance matrix $N$ is block diagonal, 
where each element of the block is $\sigma^2$, the published (noise) variance 
of that quadcube pixel. 
This means that $N$ is not strictly invertible, so we have to use a
pseudo-inverse for $N^{-1}$. 
Our pseudo-inverse is also block diagonal, with constant-valued blocks, and correctly
inverts the action of $N$ on a vector that is constant valued on the same blocks 
as $N$. 

We project out the mean and dipoles from the uncut region of the
COBE-DMR map, and model  
the data within the custom galactic cut as Gaussian random white
noise with large variance.  This corresponds to claiming complete ignorance of the
foregrounds at low galactic latitudes (within the custom cut) and
assuming that no residual foregrounds are present at high latitudes
(outside the cut region). This is the simplest possible
way of treating the monopole, dipole, and galactic foregrounds.  

Our noise matrix has values published by the COBE team, but with the
$\sigma^2$ noise variance 
increased to $1000\; mK^2$ in the galactic cut region, a numerically
large value that exceeds any other variance in the problem.

For the first iteration of the Gibbs sampler we choose 
\begin{equation}
C_0 = C_1 = 10^{-30}\; \mbox{mK}^2 \hspace{1cm}
C_\ell = \frac{10^{-4}}{\ell(\ell+1)}\; \mbox{mK}^2.
\end{equation}
We chose these values because they very roughly approximate the true $C_\ell$ values to
reduce burn-in time. The first two are numerically small, because we
consider the monopole and dipole 
to be non-cosmological.  During the $C_\ell$ estimation step of the Gibbs
sampler, the $C_0$ and $C_1$ values are not changed. This corresponds
to enforcing the prior that the cosmological signal does not contain
such components. 

The Gibbs sampler is run through $10,000$ iterations (sets of $C_\ell$ values).  
This takes approximately 24 hours on an Athlon XP1800+ workstation.
We ignore the first $1000$ iterations to ensure that the Gibbs sampler
has converged to the true distribution. This is very conservative---in
fact by computing correlations of our $C_\ell$ draws along the chain
we infer that about
every 20$^{th}$ sample is uncorrelated. 

We plot the power spectrum in
figure \ref{powspec}.   For each $\ell$ 
value, we display vertically  a binned representation of the marginalized
posterior densities $P(C_\ell|m)$. The bins all hold an equal number of points.  The bins that are thinnest 
(points are densest in $C_\ell$ space) 
are colored more darkly.  The top 68\% are dark gray; from 68\% to
95\% are lighter gray, and the rest are white. The highest density bin
is shown in black.

To explore the  marginalized posterior $C_\ell$ distribution in more
detail we plot their histograms in Figure \ref{posterior1D}. It is
noteworthy that not a single one of these is even nearly
Gaussian. Within the context of the discussion of the lack of large
scale power in the CMB, it is worth pointing out that all inferences
about $C_2$ from COBE-DMR can be based on the $P(C_2|d)$ shown here.  

The correlation structure of the estimates contains information about
how well we were able to account for the effects of the galactic
cut. It is clear from figure \ref{qMatrix} that the residual
correlations are at most of order $10\%$  even at very small
$\ell$.

However, since the posterior densities are non-Gaussian, the
two-point correlations do not contain all the 
information. We therefore show the marginalized posteriors for four pairs
of $C_\ell$s in figure \ref{posterior2D}. Again, all four of these
densities are strongly non-Gaussian. 

Lastly, we show the reconstructed signals. Figure \ref{maps}A shows
  the expectation of the  signal component (the solution of
  \eq{Wienermean} at each iteration of the Gibbs sampler) of the
  COBE-DMR data with respect to the  posterior density marginalized
  over the power spectrum: $\langle 
  x\rangle\vert_{P(x|m)}$.  This is a  generalized Wiener filter (GWF) which
  does not require knowing the signal   covariance a priori. The
  smoothing of the map autmatically adapts locally depending on how
  much detail the  data
  support. The more strongly smoothed central
  horizontal band was obscured by the galaxy. Still the GWF
  reconstructs  large scale modes in the galactic cut.

The power spectrum of figure \ref{maps}A would be biased low, since
the Wiener filter removes everything that could be noise. At each
iteration of the Gibbs sampler the solution
to \eq{Wienerfluct} (shown in figure \ref{maps}B) adds in a fluctuating term that replaces filtered
noise with synthetic signal. It is noticable that this fill-in signal
is larger in the regions of the map where the Galaxy  obscures the
CMB.
The resulting draw $s$ from \eq{signaldraw} is shown in figure
\ref{maps}C. Every $s$ draw is one possible pure signal sky that could
have given rise to the data. Since we know that the COBE data has no
statistical power above an $\ell_{max}$ of about 20, we imposed a
bandlimit of $\ell_{max}=50$.

For comparison with the inferences we draw from the COBE-DMR data, we
show in figure \ref{maps}D the internal linear combination map from
the WMAP satellite \cite{mapresults} smoothed down to five degrees
FWHM, an intermediate scale between the slightly larger average smoothing of panel A
and the somewhat smaller smoothing of panel B and C due to the
bandlimit of  $\ell_{max}=50$. Nearly every  hot and cold spot that is identified by the
GWF can be found in the high signal-to-noise WMAP data. Figure \ref{maps}C fills in signal very plausibly
up to the imposed bandlimit. Even more
striking is the similarity
of our figure \ref{maps}A to the combination of Q and V band WMAP data
shown in figure 8 of \cite{mapresults}, which is intended to mimic the
COBE-DMR 53GHz map. 

\section{Conclusions and Future Work}
\label{conclusions}
We have described a framework for global and lossless analysis of
cosmic microwave background data. This framework is based on a
Bayesian  analysis of CMB data. It has several advantages compared to
traditional methods. It is computationally feasible.  It is optimal
and exact
under the assumption of Gaussian fields and the ability to encode our
prior knowledge about foreground components in terms of multivariate
Gaussian densities. It uses
controlled approximations (e.g. the number of samples of the Gibbs
sampler controls the accuracy of the result but this can be increased
by spending more computing time). It allows
joint analysis of the CMB signal, foregrounds and noise properties of
the instrument, while modeling and exploiting the statistical dependence between these
different inferences. 

Traditional methods of inference from CMB data
divide the data analysis into several steps: map-making from TOD,
component separation, power spectrum estimation from the CMB
signal and cosmological parameter estimation. Our method allows
treating all these inferences jointly and self-consistently, if
desired. The traditional results can be understood   as special cases
of our method for certain uninformative prior choices. For example,
pure map-making could be viewed as applying this framework with $P(s|C_l) P(f)P(C_l)=const$.

In spite of this generality,  the framework for analyzing CMB data described
here is very modular: the structure of the  Gibbs sampling
scheme separates the different steps of the inference process 
focusing on each component in turn. The framework described here
therefore holds the promise of making more data analysis steps part
of a self-consistent framework rather than sequential stages in a data pipeline.

Our method turns out to give an unbiased Wiener filter and generalizes
the global filtering and reconstruction methods in \cite{rybickipress} to include power
spectrum estimation, obviating the need for a priori knowledge of the
signal covariance. 

We require the use of iterative techniques to solve  the most computationally demanding
step in this method. We find that our simple-minded preconditioned
gradient iteration
works well over 3 orders of magnitude in problem size. It remains to
be studied whether other preconditioners may be even more effective
(e.g. \cite{pen}).  

We applied our formalism to the well-studied COBE-DMR data set. We
demonstrate that our methods enable new analyses for  even
such a small data set. We quote posterior densities for
individual $C_\ell$ as well as posterior densities for pairs of $C_\ell$ as
examples of results that would be prohibitively expensive to obtain
with traditional algorithms. Our results are consistent with the most
sophisticated brute force $O(n_p^3)$ analyses available in the literature.

The approach can be  extended straightforwardly to polarized maps,  data that spans
different frequency bands and joint estimation of different data
sets. There is nothing that prevents the application of these ideas to
random fields on manifolds other than the sphere, such as one-, two-
or three-dimensional Euclidean space.
We are investigating the formalism for 
joint inference from CMB data about the power spectrum and map of the
pure CMB sky with the power spectrum and map of the projected
gravitational potential. We will report on these developments in a
future publication.

\acknowledgments
We thank I. O'Dwyer, J.~Jewell and the members of the Planck CTP
working group for comments and stimulating conversations. This work was supported in
part by the University of Illinois at Urbana-Champaign and the
NCSA/UIUC Faculty Fellowship program.

\appendix
\section{Analytic Marginalization Over  Foregrounds}
\label{marginalization}
From a statistical point of view we consider the {\it a posteriori}
distribution  to be a function of the CMB signal, $C_\ell$ and the
foregrounds. Then we marginalize over the foregrounds $f$. This can
be done either implicitly through Gibbs sampling from the joint
posterior density including $f$ (as described in the main text) and then
marginalizing over $f$ in the generated samples or explicitly
through analytic marginalization of the posterior over $f$. Then the
ignorance about the foreground is included in additional terms in the
noise covariance matrix. The first
route is more general, but there may be occasions where the second is
preferable; for example if the main goal is to make the CMB analysis
insensitive to a small number of known foreground templates $f_i$.

The effect of analytic marginalization is that the new noise
covariance matrix $N'$ including the foreground term becomes
\begin{equation}
N'\equiv N+\sigma_f^2 FF^T\equiv N+\sigma_f^2 \sum_i f_if_i^T.
\end{equation}

In order to implement the Gibbs sampler including this new term we need to be able to
apply ${N'}^{-1}$ to vectors. If only a small number (up to $\sim
1,000$) of foreground
templates need to be projected out this can best be done by grouping all the
vectors using the following limit of the
Sherman-Morrison-Woodbury formula \cite{rybickipress}
\begin{eqnarray}
N'^{-1}\equiv\lim_{\sigma_f^2\rightarrow
  \infty}(N &+& \sigma_f^2FF^T)^{-1}=\\
N^{-1}&-&\left[N^{-1}F\left(F^TN^{-1}F)^{-1}F^TN^{-1}\right) \right].\nonumber
\label{woodbury}
\end{eqnarray}
This operation will project out the directions in $N^{-1}$
corresponding to the foreground contributions. The action of this new
inverse noise covariance matrix on a vector can be computed using
methods similar to those described in Eqs.~(2.7.16{\it ff}) in
\cite{numrec}.


Alternatively one can solve iteratively the set of  equations
\begin{equation} 
\left(N+\sigma_f^2FF^T\right)x=v
\end{equation}
every time $x=N'^{-1}v$ is required on the LHS of \eq{Wienermean} and
\eq{Wienerfluct}. When $N'^{-\frac12}$ is required for the RHS of
\eq{Wienerfluct} we solve
\begin{equation} 
\left(N+\sigma_f^2FF^T\right)y=N^{\frac12}\xi+\sigma_fF\chi.
\end{equation}
The term $N^{\frac12}\xi$ is obtained by simulating a noise-only map
solving \eq{normal} with $d=n_{tod}$. In both of these equations one can
choose $\sigma_f$ numerically large.

However, the method in the main text is more flexible, since it allows
grouping different foregrounds together in ways that are
computationally convenient.

\end{document}